\documentclass{aa}  
\usepackage{graphicx}
\usepackage{txfonts}
\usepackage{lmodern}
\usepackage{amsmath}
\bibpunct{(}{)}{;}{a}{}{,}
\defcitealias{Kalvans10}{Paper~I}
\begin{document}

   \title{Modeling the processing of interstellar ices by energetic particles}

   \author{J. Kalv\=ans
          \and
          I. Shmeld
          }

   \institute{
   Engineering Research Institute
   ``Ventspils International Radio Astronomy Center''
   of Ventspils University College (VIRAC), Latvia\\
              \email{kalvans@lu.lv}
             }

   \date{Received August 6, 2012; accepted April 30, 2013}

  \abstract
   {Interstellar ice is the main form of metal species in dark molecular clouds. Experiments and observations have shown that the ice is significantly processed after the freeze-out of molecules onto grains. The processing is caused by cosmic-ray particles and cosmic-ray-induced UV photons. These transformations are included in current astrochemical models only to a very limited degree.}
   {We aim to establish a model of the ``cold'' chemistry in interstellar ices and to evaluate its general impact on the composition of interstellar ices.}
   {The ice was treated as consisting of two layers - the surface and the mantle (or subsurface) layer. Subsurface chemical processes are described with photodissociation of ice species and binary reactions on the surfaces of cavities inside the mantle. Hydrogen atoms and molecules can diffuse between the layers. We also included deuterium chemistry.}
   {The modeling results show that the content of chemically bound H is reduced in subsurface molecules by about 30 \% on average. This promotes the formation of more hydrogen-poor species in the ice. The enrichment of ice molecules with deuterium is significantly reduced by the subsurface processes. On average, it follows the gas-phase atomic D/H abundance ratio, with a delay. The delay produced by the model is on the order of several Myr.}
   {The processing of ice may place new constraints on the production of deuterated species on grains. In a mantle with a two-layer structure the upper layer (CO) should be processed substantially more intensively than the lower layer (H$_2$O). Chemical explosions in interstellar ice might not be an important process. They destroy the structure of the mantle, which forms over long timescales. Besides, ices may lack the high radical content needed for the explosions.}

   \titlerunning{Modeling of interstellar ices}
   \maketitle

\section{Introduction}
\label{intro}

One of the important periods in the transformation process of gas into stars is the pre-stellar core (PSC) phase. This period lasts not much longer than one Myr \citep{Lee99, Enoch08, Schnee12}. Most elements heavier than He (referred to throughout as metals) are either locked in the interstellar dust-grains, or freeze onto the grains as icy mixtures.

The ice is formed by the accretion of atoms and the subsequent formation of molecules on the surface, such as H$_2$O and most other species, or by the direct accretion of molecules, such as CO \citep{Oberg11-109}. The formation occurs first in steps that affect the chemical composition and physical structure of the ice. In the dark PSCs it is processed mainly by cosmic rays (CR) and cosmic-ray-induced photons as well as interstellar ultraviolet (UV) and X-ray photons, radioactive decay events, and heating by exothermic reactions. Particularly important are heavy cosmic rays, mostly Fe nuclei (Fe-CR) \citep{Leger85}.

Astrochemical models usually include gas and grain surface as the chemically active phases \citep[e.g.][]{Hasegawa93b, Roberts07, Albertsson11}. However, both observations of interstellar matter and astrochemistry laboratory simulations have shown that the processing of ice is not limited to the outer surface.

First, the interstellar ice seems to be compact instead of porous. The porous structure is expected to form from molecules accreting on the surface at 10K. The timescale for mantle compaction calculated from experiments with UV photons, exothermic reactions, and energetic ions is from a few up to 50 Myr \citep{Raut08, Palumbo10, Accolla11}. The icy mantles tend to be mixtures. Their compaction is slower than that of pure substances and it could never be completed during the cloud lifetime \citep{Palumbo06}. Observations of pure ices and mixtures consisting of two species indicate that the mantle is segregated in chemically different layers \citep{Pontoppidan03, Oberg11-109, Linnartz11}. The segregation can be attributed to the heating for ices already in the protostellar phase \citep{Oberg09a}. Temporal heating events by UV photons and cosmic rays may induce ice segregation in prestellar phases. Among other things, the observation of differentiated layers indicates the ability of molecules to move along the thickness of the ice over timescales relevant to PSCs.

Second, the ice mantle on the grains is subjected to chemical processing by UV photons and cosmic rays. Multiple experiments \citep[e.g.][among recent papers]{Gerakines96, Gerakines04, Andersson08, Oberg09a, Oberg09b, Oberg11n} show that the composition of ice irradiated by ionizing photons can be quite different from the initial one, regardless of whether it was pure substance or a mixture. Similarly, suprathermal reactions and the appearance of new species are induced by ice irradiation by fast ions \citep[e.g.][Chap.~14]{Millar93}. Mechanisms for these processes include the generation of radicals, migration, and reactions of species in elevated temperature.

A solution was proposed to quantitatively describe the ``cold'' processing of the mantle by \citet[][Paper I]{Kalvans10}. The model described there includes a few key concepts:
\begin{itemize}
      \item The molecules frozen onto the grains are assumed to be divided into two layers -- the surface and the mantle. The mantle also includes molecules on the surfaces of closed cavities inside the ice. Thus, there is a total of three phases -- the outer surface, the mantle volume, and the cavities.
			\item Molecules in these phases are subjected to photodissociation by UV photons. In the dark conditions of PSCs, only CR-induced photons have a sufficiently high flux to be important.
			\item Reactions occur on the outer surface and the surface of the cavities. The inclusion of the chemically active cavity phase makes the system different from the three-phase model of \citet{Hasegawa93b}. The reactions in the cavities consume the radicals generated in mantle by photodissociation.
			\item The cavities themselves are processed (resurfaced, destroyed, generated again) by cosmic-ray hits. Most of the molecules and radicals in the mantle are cycled through the cavity phase and thus the reactions on the inner surfaces are able to affect the whole mantle.
			\item The diffusion of hydrogen occurs through the mantle. This process is described as a transition from one ice phase to another. Only surfaces are considered as the stable residence sites for H and H$_2$, following \citet{Strauss94}. In \citetalias{Kalvans10} only the outer and cavity surfaces were considered for the diffusion of H. In this work (Sect.~\ref{Hdiff}) the mantle volume phase is included, too. It is represented by micropores that can host one or several H atoms, but are too small for reactions involving metal atoms to occur.
\end{itemize}

Notably, the model does not include chemical transformations induced directly by cosmic-ray hits. The astrochemical database used (udfa06) does not contain dissociation reactions for metal molecules hit by cosmic-ray protons.

In this paper we aim to study the effects of subsurface mantle chemistry in more detail. The approach outlined in \citetalias{Kalvans10} is kept with some improvements. The time-dependence for the evolution of the composition of the mantle and deuterium chemistry are investigated. Molecules with a high enrichment of deuterium are characteristic for PSCs \citep[see, e.g.][]{Caselli11}. There are indications that the solid species of the icy grain-mantles are highly deuterated, too, and evaporate to the gas-phase \citep[e.g.][]{Mangum91}. The deuteration is usually attributed to gas phase chemistry with the isotopologs of  $\mathrm{H_3^+}$ \citep{Millar89, Roberts00, Roberts04} that act as important intermediate reactants. They produce an overabundance of atomic D, which then accretes onto dust-grains and reacts with other surface molecules, producing highly deuterated species. When the accretion continues, surface is covered and surface molecules are incorporated into the mantle. In our model they are still subjected to the dissociation by CR-induced UV photons and chemical reactions on cavity surfaces.

The paper is organized as follows. General chemical input data, information on the gas-phase chemistry, and specific information about photoreactions are given in Sect.~\ref{chem}. All processes that are directly related to surfaces are described in Sect.~\ref{surf}. The most important novelty introduced in the model is the concept of cavities within the ice mantle. The processes related specifically to the cavities are described in Sect.~\ref{transf}. At the end of the description of the calculation methodology the summary rate equations are given in Sect.~\ref{sums}. The outcome of the simulation is described in detail in Sect.~\ref{results}. Finally, the conclusions form the results are briefly discussed in Sect.~\ref{disc}.

\section{Chemical data}
\label{chem}

A cloud with constant density of hydrogen atoms $n_{\mathrm{H}}=10^{5}$cm$^{-3}$, gas temperature $T_g = 15$ K, and dust temperature $T_d =10$ K was modeled for the integration times up to $t=9.5\times10^{13}$ s, which equals approximately 3 Myr. The integration times used are beginning from $1.4\times10^{12}$ s, multiplied by a factor of 1.42 and rounded to two significant numbers until a total of 13 steps was reached. Initially, all elements were assumed to be in atomic form. Calculations with shorter integration times were considered to be scientifically irrelevant.

\subsection{Species and reactions}
\label{specs}

The UMIST udfa06 dipole astrochemistry database \citep{Woodall07} was used to provide the gas-phase chemical reaction set (Sect.~\ref{gas}). The permitted elements were H, D, He, C, N, O, Na, Mg, S, and Fe, and their relative abundance and depletion are specified in Table~\ref{tab-elements}.

In total, the model includes 450 chemical species with more than 10,000 reactions. No species with more than four heavy atoms and four hydrogen atoms are included, with the sole exception of CH$_{5}^+$ and its hydrogen isotopologs. Of these species, 186 neutral species are refractory, accreting and participating in the chemistry on the grains. We reduced the number of chemical species to facilitate the calculations. The aim of this paper is to investigate the general effects of the subsurface processes, not to model the whole chemistry of an interstellar cloud core.

Outer and cavity surface reactions are described with the approach described by \citet{Hasegawa92} (Sect.~\ref{surf-reac}). The dissociation by CR-induced photons for all ice species is included, with the rates provided in the gas-phase udfa06 database and modified by the respective dissociation yield for each ice phase (Sect.~\ref{CRphdis}).
\begin{table}
\caption{Adopted abundances and depletion of elements in the model.}
\label{tab-elements}
	\centering
	\small
		\begin{tabular}{ccccc}
\hline\hline
 &Total$^{\mathrm{a}}$& & &Initial$^{\mathrm{a}}$\\
Element&abundance&Depletion,&Source&gas-phase\\
 & & \% & &abundance\\
\hline
H&1.00&0&--&1.00\\
D&2.00E-05&0&PSF10$^{\mathrm{b}}$&2.00E-05\\
He&0.10&0&--&0.10\\
C&2.88E-04&0.39&J09$^{\mathrm{c}}$&1.77E-04\\
N&7.94E-05&0.22&J09&6.18E-05\\
O&5.75E-04&0.42&J09&3.35E-04\\
Mg&4.17E-05&0.95&J09&2.25E-06\\
Fe&3.47E-05&0.99&J09&2.01E-07\\
S&2.10E-05&0.92&J09, P97$^{\mathrm{d}}$&1.77E-06\\
Na&2.14E-06&0.75&P97&5.34E-07\\
&&&PPG84$^{\mathrm{e}}$&\\
\hline
		\end{tabular}
\footnotesize
\begin{list}{}{}
\item[$^{\mathrm{a}}$] relative to hydrogen
\item[$^{\mathrm{a}}$] \citet{Prodanovic10}
\item[$^{\mathrm{b}}$] \citet{Jenkins09}, with $F_*=1$
\item[$^{\mathrm{c}}$] \citet{Pagel97}
\item[$^{\mathrm{d}}$] \citet{Phillips84}
\end{list}
\end{table}

All processes in the model are described in terms of rate equations. This applies to first and second order chemical reactions and phase transitions  (Sects.~\ref{gas},~\ref{CRphdis},~\ref{accr},~\ref{desorp},~\ref{transf}). The general rate equation for a first-order process is
   \begin{equation}
   \label{rate1}
\frac{\mathrm{d}n_{i,f}}{\mathrm{d}t} = k_1 n_{i,f0},
   \end{equation}
where d$n_{i,f}$ is the abundance change (cm$^{-3}$s$^{-1}$) of species \textit{i} in phase \textit{f}, $k_1$ is the rate coefficient (s$^{-1}$), and $n_{i,f0}$ is the abundance of \textit{i} in the initial phase (cm$^{-3}$). The phases \textit{f} and \textit{f0} are either gas \textit{g}, outer-surface \textit{S}, cavity-surface \textit{C}, or mantle \textit{M}. For chemical reactions (usually) $f=f0$.

The rate (cm$^{-3}$s$^{-1}$) of second-order reactions of type

$l + j \rightarrow i$

is calculated by the equation
   \begin{equation}
   \label{rate2}
\frac{\mathrm{d}n_{i,f}}{\mathrm{d}t} = k_{lj,f}n_{l,f}n_{j,f},
   \end{equation}
where $k_{lj,f}$ (cm$^{3}$s$^{-1}$) is the second-order rate coefficient.

\subsection{Including H isotopolog reactions}
\label{Dchem}

Deuterium is involved in the reaction network following the approach by \citet{Rodgers96}. That is, the rate of a gas-phase reaction is equal for deuterated and non-deuterated species and statistical branching ratios for reaction outcomes are assumed. For surface reactions, the reaction rate of deuterium species is modified to take into account their different molar mass according to \citet{Hasegawa92}. An approximation is made that in a molecule there is no distinction between H or D atoms attached to different heavy atoms. For example, methanol is treated as CH$_4$O, not CH$_3$OH. This approximation facilitates calculations at the cost of only little loss of information because the actual fate of different hydrogen atoms in a reaction is usually unknown anyway. The gas-phase deuterium exchange reactions given by \citet{Roberts00} and \citet{Roberts04} are added to the reaction list.

\subsection{Gas-phase reactions}
\label{gas}

The rate of first-order reactions in gas phase ($f=f0=g$) is calculated according to Eq.~\ref{rate1} with the rate coefficient $k_{udfa1}$ (s$^{-1}$) from the udfa06 database. It is obtained as explicitly given in \citet{Woodall07}, Eqs. (2), (3), and (4). The first-order reactions are direct cosmic-ray ionization, cosmic-ray-induced photoreactions, and interstellar photoreactions. For the latter the extinction of the interstellar UV radiation field $A_V=20$ mag, which makes these photoreactions rather unimportant.

The rate of second-order gas-phase reactions is calculated in accordance with Eq.~\ref{rate2} ($f=g$). The respective rate constant $k_{udfa2}$ is obtained by the reaction coefficients given in udfa06, as in \citet{Woodall07}, Eq. (1).

\subsection{Solid-phase cosmic-ray-induced photodissociation}
\label{CRphdis}

The dissociation by cosmic-ray-induced UV photons for gas, surface, mantle, and cavity phase molecules is based on data from the udfa06 database. The photodissociation rate for a solid-phase molecule ($f=f0=S,C$ or \textit{M}) \textit{i} is a first-order process, calculated according to Eq.~\ref{rate1} with a rate coefficient
   \begin{equation}
   \label{CRphdis2}
k_{ph.dis,f} = k_{CRPHOT}Y_{dis,tot,f}.
   \end{equation}
$k_{CRPHOT}$ is the respective rate coefficient from the udfa06 database (s$^{-1}$) and $Y_{dis,tot,f}$ is the quantum yield for dissociation in the solid phases. We assumed that the dissociation properties for molecules in ice remain the same as in the gas-phase. Ionization is not taken into account for solid-phase species. $Y_{dis,tot,f}$ is different for each of the three ice phases and is calculated by 
   \begin{equation}
   \label{CRphdis3}
Y_{dis,tot,f} = Y_{ph.trav,f}Y_{dis,f}.
   \end{equation}
$Y_{ph.trav,f}$ is the probability for a photon to reach a monolayer of certain (average) depth in the mantle, characteristic for the phase \textit{f}. $Y_{dis,f}$ is the probability that the dissociation products survive in the phase \textit{f} and do not recombine \textit{in situ}. For mantle- and cavity-phase molecules $Y_{ph.trav,f}$ corresponds to the middle of a 100-monolayer thick mantle. The absorption in UV for each monolayer is 0.02 \citep{Hartquist90}.
\begin{table}
\caption{Adopted values for cosmic-ray-induced photodissociation yields and probabilities for hydrogen diffusion (Sect.~\ref{Hdiff}) for the three ice phases.}
\label{tab-phases}
	\centering
		\begin{tabular}{ccccc}
\hline\hline
Phase (\textit{f})&$Y_{ph.trav,f}$&$Y_{dis,f}$&$Y_{dis,tot,f}$&$P_{diff,f}$\\
\hline
Surface (\textit{S})&1.0&0.5&0.5&0.5\\
Cavities (\textit{C})&0.4&0.5&0.2&0.1\\
Mantle (\textit{M})&0.4&0.01&0.004&0.01\\
\hline
		\end{tabular}
\end{table}

$Y_{dis,f}$ is largely unknown, especially for astronomically relevant timescales, although many photolysis experiments have been conducted (see Sect.~\ref{intro}). We used values that produce feasible results and seem physically adequate. Higher photodissociation yields need a higher proportion  of ice species in the chemically active cavity phase $X_{cav}$, so that the radicals can be consumed and turned into molecules. $Y_{dis,f}$ is adjusted for $X_{cav} \approx 1\%$ with a radical content of a few per cent at most. On the outer surface, dissociation products H and $\mathrm{H_2}$ and their deuterium analogs escape into the gas phase as described in \citet{Hartquist90} \citepalias[see also][]{Kalvans10}.

In addition to the three solid phases, photodissociation is separately considered for mantle-phase molecules that happen to reside directly next to a cavity. The proportion in ice of these species is assumed to be equal to $X_{cav}$, i.e., 1 \%. In the model their dissociation products enter the cavity and thus change their phase ($f=C$ and $f0=M$). The dissociation yield for these molecules is the same as for other mantle-phase species, and the rate coefficient (s$^{-1}$) is
   \begin{equation}
   \label{CRphdis4}
k_{ph.dis,MC} = 0.01k_{CRPHOT}Y_{dis,tot,M}.
   \end{equation}
\section{Surface-related processes}
\label{surf}

\subsection{Accretion}
\label{accr}

The outer surface is defined as the porous layer of molecules on top of the denser mantle. Surface-phase species are exposed to surface reactions and desorption processes. The rates of neutral molecule and atom accretion onto grains are calculated according to the first-order rate equation~\ref{rate1} with $f=S$ and $f0=g$. The relevant rate coefficient (s$^{-1}$) is calculated according to \citet{Willacy93}:
   \begin{equation}
   \label{accr2}
k_{accr} = 3.2\times10^{-17}n_{\mathrm{H}}S_{i}\left(\frac{T_{g}}{\mathrm{M}_{i}}\right)^{1/2},
	 \end{equation}
where $S_{i}$ is the assumed sticking coefficient, M$_i$ is the molecular mass of species \textit{i} in atomic mass units, and $n_\mathrm{H}$ is the total hydrogen nucleon density. Following \citet{Brown90}, $S_{metal}=1.0$. For the light species H, D, $\mathrm{H_2}$, HD, and $\mathrm{D_2}$ a lower value is used, $S_{\mathrm{H}}=1/3 S_{metal}$, as in \citetalias{Kalvans10}, i.e. $S_{\mathrm{H}}=0.33$.

\subsection{Desorption}
\label{desorp}

The desorption rate of species from the surface into the gas-phase ($f=g, f0=S$) is a first-order process, Eq.~\ref{rate1}. $k_{des}$, s$^{-1}$, is the rate coefficient. There are five desorption mechanisms included in the model. For thermal evaporation the rate coefficient is the inverse evaporation time $t_{evap}$ from \citet{Hasegawa92}, Table 1. The desorption via grain heating by cosmic rays is treated as in \citet{Hasegawa93a}:
   \begin{equation}
   \label{des12}
k_{crd}=f(70\mathrm K)k_{evap,i}(70\mathrm K),
   \end{equation}
where \textit{f}(70K) is the fraction of the time spent by grains in the vicinity of 70K and $k_{evap,i}$(70K) is the evaporation rate coefficient for species \textit{i} at 70K \citep[, Eq.~14]{Hasegawa93a}. \textit{f}(70K) is calculated as
   \begin{equation}
   \label{des13}
f(70\mathrm K)=\frac{t_{cool,70K}}{t_{CR}}.
   \end{equation}
$t_{cool}$, s, is the time-scale for grain cooling via desorption of volatiles (10$^{-5}$s) and $t_{CR}$ is the average time between two successive Fe-CR hits on the grain. $t_{CR}=10^{12}$s from \citet{Bringa04} results in a rate coefficient higher than that in \citet{Hasegawa93a}.

The rate coefficient for desorption by CR-induced photons \citep{Prasad83, Willacy93} is calculated by
   \begin{equation}
   \label{des2}
k_{crpd}=R_{ph}Y,
   \end{equation}
where the photon hit rate $R_{ph}$ (s$^{-1}$) for a single grain is
   \begin{equation}
   \label{des3}
R_{ph}=F_{p}\left\langle \pi\alpha^{2}_{g}\right\rangle.
   \end{equation}
The desorption yield \textit{Y} is taken as 0.1, $F_{p}$ is photon flux, 4875 cm$^{-2}s^{-1}$, both from \citet{Roberts07}. $\left\langle \pi\alpha^{2}_{g}\right\rangle$ is the average cross section of a grain ($1.0\times10^{-10}$ cm$^{2}$).

The rate of desorption of surface molecules by energy released by $\mathrm{H_{2}}$ molecule formation on grains is calculated according to the rate coefficient \citep{Roberts07}:
   \begin{equation}
   \label{des4}
k_{\mathrm{H_2}f.des}=3.16 \times 10^{-17} \epsilon n(\mathrm{H}) n_\mathrm{H},
   \end{equation}
where $\epsilon=0.01$ is number of molecules desorbed per act of H$_{2}$ molecule formation, \textit{n}(H) is the number density of atomic hydrogen in gas-phase \citep{Roberts07}. The species desorbed are those with binding energies $E_{b} \leq 1210$K, with binding energies from \citet{Aikawa97}.

Summarizing the desorption rate coefficients:
   \begin{equation}
   \label{des7}
k_{des} = t^{-1}_{evap} + k_{crd} + k_{crpd} + k_{\mathrm{H_2}f.des}.
   \end{equation}

In addition to desorption from the surface, we included a process of direct impulsive ejection of \textit{mantle} species into the gas \citep{Johnson91} ($f=g, f0=M$). The rate coefficient (s$^{-1}$) is calculated as in \citetalias{Kalvans10}:
   \begin{equation}
   \label{des6}
k_{ej} = t^{-1}_{CR} Y_{CR}.
   \end{equation}
$Y_{CR}$ is the number of mantle molecules ejected by this process in each hit, assumed to be 1000 here, close to the upper limit from \citet{Johnson91}. It is the only direct mechanism for molecules to escape the subsurface layers of the frozen ice on grains. Impulsive ejection for surface species is insignificant compared to the desorption rates for other mechanisms and is not included in the model.

\subsection{Binary reactions on surfaces}
\label{surf-reac}

The rate of binary reactions (second-order process, cm$^{-3}$s$^{-1}$) on the outer surface ($f=S$) and the cavity surface ($f=C$) is calculated according to Eq.~\ref{rate2}. The respective rate coefficient is calculated as in \citet{Hasegawa92}:
   \begin{equation}
   \label{surf2}
k_{lj,S/C}=\frac{\kappa_{lj} (R_{mov,l} + R_{mov,j}}{n_d}).
   \end{equation}
$R_{mov}$ is the movement (diffusion on surface) rate for each respective reactant, $\kappa_{lj}$ is the probability for molecules to overcome the activation energy barrier upon encounter for the reaction to take place \citep[Eq.(6) in][]{Hasegawa92}, and $n_d=1.33 \times 10^{-12} n_\mathrm{H}$ cm$^{-3}$ is the abundance of dust-grains. $R_{mov}$ is given by
   \begin{equation}
   \label{surf3}
R_{mov}=N_{s,f}^{-1} t_{hop}^{-1},
   \end{equation}
where $N_{s,f}$ is the number of adsorption sites on one of the two chemically active surface types (phases), $t_{hop}$ is the hopping time from one adsorption site to the next for a particular molecule, calculated as given by \citet{Hasegawa92}, Eqs. (2), (3) and (10). For deuterium-containing species the movement rates are recalculated taking into account their increased molecular mass.

\citet{Hasegawa92} used a value $N_{s,S}=10^6$ for their surface reaction rate calculations. An increased value of $N_{s,S}=2 \times 10^6$ fits this simulation better, because surface roughness is an important feature of the model. The abundance of metal atoms in outer-surface and cavity phases is similar in the final integration time steps (Sect.~\ref{res-str}), so $N_{s,C}$ can be taken to be similar to $N_{s,S}$. $N_{s,C}=10^6$, because the abundance of cavity species is significantly below the abundance of surface species and molecules will have fewer options for adsorption in the cramped cavities. For a simple representation one can imagine a mantle consisting of 100 monolayers that includes 1000 cavities that have 1000 adsorption sites each. It is important to note that $N_{s,C}$ and $N_{s,S}$ are assumed to be constant in the model, although the relative abundance of metals in surface- and mantle phases changes with time (Sects.~\ref{mantle} and~\ref{res-str}).

\subsection{Formation of the mantle}
\label{mantle}

The mantle is formed by the transfer of molecules from the surface- to the mantle-phase. It is described as a reversible first-order phase transition, Eq.~\ref{rate1} with $f=M$ and $f0=S$ for the forward and $f=S$ and $f0=M$ for the reverse process. The gradual compaction of the ice layer is described by this approach in the model (see below).

Currently the structure of the ice phase on grains is usually described by two different methods in astrochemical modeling . Both are capable of yielding feasible results. Subsurface chemistry is not taken into account by any means in either of the calculation methods.

One method is to describe the ice mantle as a rigid structure, as e.g. in \citet{Hasegawa93b} or \citet{Cuppen09}. The surface species are mobile, reactive, and subjected to desorption. Molecules in the layers below are frozen in place, forming a densely packed permanent structure divided into monolayers.

However, repeated experiments have shown that ices formed at cryogenic temperatures are porous \citep{Accolla11, Palumbo10, Raut07a, Raut08}. Molecules of considerable size (CO, CH$_4$) are able to diffuse through an accumulate in these interstellar ice analogs \citep{Palumbo06, Raut07b}. These molecules are initially deposited on the surface. Their diffusion into the depth of the ice layer means that the molecules in the porous ice are available for surface reactions. Moreover, the compaction of ice (see below) indicates that the molecules are able to move within the mantle in conditions and timescales relevant to dark clouds. One can conclude that a rigid mantle is a rough approximation of the astrophysical reality.

A simpler way is to treat the ice as a single phase, consisting of reactive (surface) species \citep[e.g.][]{Roberts07, Semenov10, Weeren09}. In terms of this model it can be described as a porous layer, fully exposed to surface reactions. The experiments mentioned above have also shown that the porous ice becomes compact when subjected to energetic processes (ion and UV-photon irradiation, exothermic reactions). Observations of interstellar ices indicate that they are dense and segregated in chemically different layers \citep[e.g.][]{Jenniskens95, Oberg09b}. The time of compaction is estimated to be from a few to 50 Myr, depending on the process considered. Icy mixtures, characteristic of interstellar ices, may be never completely compacted \citep{Palumbo06}.

When depleted metal abundances are used, one can assume that the single-ice-phase model includes the very upper layer of the ice mantle only. The surface phase is intimately connected to the gas phase. The model contains no direct information about the composition of the bulk of the ice.

We propose a simple way to include the behavior of interstellar ices in the model, taking into account the experimental findings. It is reasonable to assume that the time of ice compaction (Sect.~\ref{intro}) is longer than the time of accretion of molecules onto grains in dense cores \citep[$10^5$ years,][]{Schutte91}. As a basic assumption, the accretion and ice compaction does not occur simultaneously. More than 90 \% of the metal atoms are already present in the ice phases in the first integration time-step of the simulation.

The calculation of the transition rate of species into the mantle phase is related to the rate of heavy cosmic-ray hits on the grain. In our view light CRs, UV, and X-ray photons, exothermic reactions, and other events all contribute to the compaction of ice. The rate of this first-order process is calculated according to Eq.~\ref{rate1}, with $f=M$ and $f0=S$. The rate coefficient (s$^{-1}$) is
   \begin{equation}
   \label{mant3}
k_{mantle} = t^{-1}_{CR} v^{-1},
   \end{equation}
where \textit{v} is the number of Fe-CR strikes, during which all surface species are converted into mantle species.

The percentage of outer-surface species in completely compacted ice with a thickness of 100 monolayers can never be lower than 1 \%. A reverse process - the transition of ice species from the dense mantle-phase to the porous and exposed surface-phase ($f=S, f0=M$) helps to describe the formation and compaction of the ice. The respective rate coefficient is $k_{dig-up}$. To have approximately 1\% of ice-forming species in the outer-surface phase, $k_{mantle}$ is assumed $100k_{dig-up}$.

\textit{v}=50 and $k_{dig-up}:k_{mantle}=1:100$ results in a slowly growing mantle-to-surface metal atom abundance ratio up to approximately 100:1. After $~5\times10^{13}$s it remains roughly constant (Sect.~\ref{res-str}). This simulates the gradual compaction of the ice.

The assumption of molecule transition from the mantle- to the surface phase is explained as follows. Cosmic-ray hits and other energetic events induce molecule movement within the ice that leads to compaction and segregation \citep{Oberg09a}. As a side effect, some mixing occurs, and some molecules in the denser mantle become exposed to the outer surface or surface of open pores. However, the full mixing of the mantle is unlikely, and we regard the transition of mantle molecules to the surface as an approximation, only. Because the timescales of mantle circulation are much longer than those for other processes, this approximation can be considered tolerable for the purposes of the model. If the ice is initially porous in interstellar conditions, this model might be conceptually closer to the physical reality than most other current models. A low-level porosity of the mantle in the model is always maintained via the concept of the cavity phase (next section).

The generalized approach to modeling the mantle porosity and cavities is a tool for simulating the ice chemistry. The structure of the mantle is described essentially with a set of input parameters. This is different the work by \citet{Cuppen07} and \citet{Cuppen09}, where detailed simulation of the formation and structure of the ice layer has been performed using the continuous-time random-walk Monte Carlo simulation technique.

\section{System of cavity-related transformations}
\label{transf}
   \begin{figure}
   \vspace{1.5cm}
   \includegraphics[width=9.0cm]{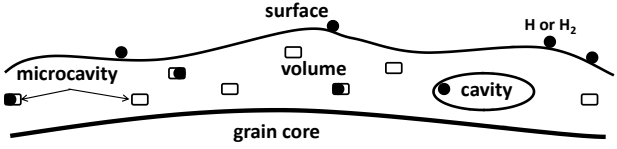}
   \vspace{-5.0cm}
   \caption{Schematic drawing of the mantle, as reflected in the model. The ice phases - surface, cavities, and mantle volume, which includes micropores, are shown. Not to scale.}
   \label{att-mantle}
   \end{figure}
In Section ~\ref{intro} (this paper) and in Paper I, we have described the model that reflects the processing of the mantle.  Fig.~\ref{att-mantle} shows a sketch of the present mantle model. There are improvements over Paper I, because hydrogen diffusion to micropores (see Sect.~\ref{Hdiff} below) and CR-induced photoreactions of mantle-phase species (Sect.~\ref{CRphdis}) are included in this model. For short, the mantle always has some degree of porosity and its molecules are subjected to photodissociation, which produces chemically active radicals. The lighter radicals can move across the surfaces of pores and react when an appropriate reactant is found. In this way, they are converted into stable molecules again. Meanwhile, hydrogen diffuses between the mantle and the outer surface. That means, H and H$_2$ released by dissociation may escape the mantle. They may also diffuse into the mantle from the surface, which is exposed to the gas phase. The calculations show that these two effects do not exactly balance out.

\subsection{Formation of the cavities}
\label{cavs}

We assumed that molecules are periodically cycled through the chemically active cavity phase by movement induced by heating from Fe-CR hits and other energetic events. The rate of molecule transition between the cavity phase and the mantle phase is described as a first-order phase transition, Eq.~\ref{rate1}. The respective rate coefficients are $k_{act}$ for $f=C, f0=M$ and $k_{inact}$ for $f=M, f0=C$:
   \begin{equation}
   \label{cav3}
k_{act}=t^{-1}_{CR}\times u^{-1}
   \end{equation}
and
   \begin{equation}
   \label{cav4}
k_{inact}=t^{-1}_{CR},
   \end{equation}
where \textit{u} is the number of Fe-CR strikes during which the grain mantle is fully reorganized. That means, the amount of molecules exposed to a cavity surface is equivalent to the total number of molecules per mantle. \textit{u} is assumed 100 (see below) and all existing cavity surfaces are lost (resurfaced) upon the next Fe-CR hit (Eq.~\ref{cav4}). The activation of intact molecules by cosmic-ray hits and dissociation by CR-induced UV radiation of molecules nearby to the cavities (Eq.~\ref{CRphdis4}) are the two processes in the model that change the molecules exposed in the cavities in the mantle.

The rate coefficients of mantle-cavity phase transitions are attached to the rate of Fe-CR hits in a way similar to the case of mantle formation. The ratio $k_{act}/k_{inact}$ = 0.01 determines the proportion of mantle species in the cavity phase ($X_{cav}$). It is assumed that the continuous bombardment by high-energy particles always keep 1\% of subsurface mantle species on the surfaces of large, closed pores, channels, or cracks. They are considered chemically active. In our opinion it is close to the maximum percentage that is still consistent with the view that interstellar ices are amorphous and dense \citep{Raut08, Accolla11, Palumbo06}. It ensures sufficiently rapid consumption of radicals in the mantle, although their content is still considerable at 1\%.

\subsection{Hydrogen diffusion through the mantle}
\label{Hdiff}

For hydrogen diffusion each of the three solid phases plays an important role. According to the conclusions by \citet{Strauss94}, hydrogen almost always resides on a surface, and H and H$_2$ residing in the lattice is only a rare intermediate state. The model assumptions are as follows. First, hydrogen atoms and molecules accrete onto the grain (mantle) outer surface. They also reside on inner surfaces in the cavity phase and in the more densely packed amorphous volume phase - in microcavities. A microcavity is too small to provide enough surface and reactants for chemical reactions, but it is large enough to provide a stable surface for one or a few hydrogen atoms/molecules to reside on. Thus the two mantle phases also represent two kinds of cavities, when regarding the diffusion of hydrogen in the model. This two-level division of pores in interstellar ices is consistent with the experimental findings by \citet{Raut07b}. The rate of transition for H and H$_2$ from solid-phase $f_0$ to solid-phase \textit{f} is calculated as a first-order process by Eq.~\ref{rate1}. The respective rate coefficient is
   \begin{equation}
   \label{hdiff2}
k_{diff,f}=P_{diff,f} \frac{4D}{L^2},
   \end{equation}
where \textit{D} is the diffusion coefficient (cm$^2$s$^{-1}$), $P_{diff}$ is the probability for diffusing species to find an appropriate surface of the phase \textit{f}, hereafter referred to as \textit{P} for short. $L$ is the length of the diffusion path, $4\times10^{-6}$cm, approximately half the thickness of the mantle. $D_{\mathrm{H}}=2.5\times10^{-21}$cm$^2$s$^{-1}$ from \citet{Awad05} and $D_{\mathrm{H_2}}=5.9\times10^{-8}$cm$^2$s$^{-1}$, as estimated from the data by \citet{Strauss94} in \citetalias{Kalvans10}.

The diffusion coefficient of deuterium species through the icy mantle is very poorly known. In fact, no usable calculation or measurement data could be found on \textit{D} for D, HD, and $\mathrm{D_2}$ at cryogenic temperatures in an environment similar to interstellar ices. We estimated the coefficients from the available literature on the diffusion of deuterium species in cryogenic temperatures \citep{Guil73, Kappesser96, Martin96, Forsythe98, Kua01, Coulomb03}. For the sake of convenience \textit{D} for D and HD, D$_2$ is expressed as $D_\mathrm{H}$ or $D_{\mathrm{H_2}}$, respectively, modified by a multiplier. The multipliers for D, HD and D$_2$ are 10$^{-7}$, 10$^{-6}$ and 10$^{-8}$, respectively. In our opinion these values could be at the lower-end of plausible values for \textit{D}, i.e., meaning rather slow diffusion.

\begin{figure}
	 \vspace{1.0cm}
	 \hspace{0.5cm}
   \includegraphics[width=7.0cm]{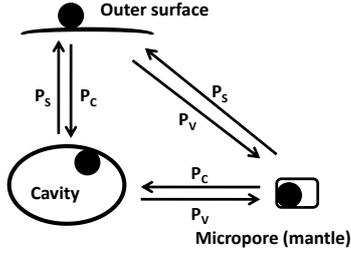}
   \vspace{-2.5cm}
   \caption{Diffusion scheme of H, H$_2$ and their isotopologs. $P_S$: the probability for diffusion to the outer-surface phase, i.e., $f=S$, $P_C$: $f=C$, $P_M$: $f=M$.}
   \label{att-hdiff}
   \end{figure}

The hydrogen diffusion probabilities to the outer surface $P_S$, to cavities $P_C$, and to the volume of the mantle (microcavities) $P_M$ are dependent on their relative surface area and the number of available adsorption sites. They are summarized in Table~\ref{tab-phases}. Fig.~\ref{att-hdiff} shows the respective probabilities for diffusion to each of the ice phases.

\section{Rate equations}
\label{sums}

The actual rate of abundance change is calculated by summing the rate of the first-order phase transitions $i(f0) \rightarrow i(f)$, first-order reactions $j \rightarrow i$, second-order reactions $l + j \rightarrow i$, and $i + j$. The total rate of change for the abundance of gas-phase species \textit{i} can be expressed as
	\begin{multline}
   \label{rateg}
\frac{\mathrm{d}n_{i,g}}{\mathrm{d}t} = \sum^{}_{lj}k_{lj,g}n_{l,g}n_{j,g}- n_{i,g}\sum^{}_{i \neq j}k_{ij,g}n_{i,g} \\
- k_{accr,i}n_{i,g} + k_{des,i}n_{i,S} + k_{ej}n_{i,M}.
	\end{multline}
For the outer-surface phase species total rate of abundance change is
	\begin{multline}
   \label{rateS}
\frac{\mathrm{d}n_{i,S}}{\mathrm{d}t} = \sum^{}_{lj}k_{lj,S}n_{l,S}n_{j,S}- n_{i,S}\sum^{}_{i \neq j}k_{ij,S}n_{i,S} \\
+ k_{accr,i}n_{i,g} - k_{des,i}n_{i,S} \\
- k_{mantle}n_{i,S} + k_{dig-up}n_{i,M}.
	\end{multline}
For the cavity-surface phase species:
	\begin{multline}
   \label{rateC}
\frac{\mathrm{d}n_{i,C}}{\mathrm{d}t} = \sum^{}_{lj}k_{lj,C}n_{l,C}n_{j,C} - n_{i,C}\sum^{}_{i \neq j}k_{ij,C}n_{i,C} \\
+ k_{act,i}n_{i,M} - k_{inact,i}n_{i,C} + 0.01k_{j,dis,tot,M}n_{j,M},
	\end{multline}
and for the mantle-phase species:
	\begin{multline}
   \label{rateM}
\frac{\mathrm{d}n_{i,M}}{\mathrm{d}t} = \sum^{}_{i \neq j}k_{j,dis,tot,M}n_{j,M} - 1.01k_{i,dis,tot,M}n_{i,M} \\
+ k_{mantle}n_{i,S} - k_{dig-up}n_{i,M} \\
- k_{act,i}n_{i,M} + k_{inact,i}n_{i,C}.
	\end{multline}
For the light hydrogen species H, H$_2$, D, HD, and D$_2$ the ice diffusion terms (Sect.~\ref{Hdiff}) have to be added for rate calculation in \textit{S}, \textit{C}, and \textit{M} phases.

\section{Calculation results}
\label{results}

\subsection{Method for comparing of abundances in the ice phases}
\label{ms}

We here and in \citetalias{Kalvans10} investigate for the first time the chemical transformation of subsurface ice in the mantles of interstellar dust grains. The most important task is to study the changes in molecular abundances introduced by the subsurface chemistry relative to outer-surface abundances. To compare this, the ratio between the mantle- and surface phase relative abundances for species can be used (mantle-to-surface ratio, \textit{M/S}),
\begin{equation}
   \label{ms1}
M/S_i=\frac{n_{i,M}}{n_{i,S}}.
   \end{equation}
However, as described in \citetalias{Kalvans10}, the best tool for demonstrating the relative changes in mantle and surface abundances is the normalized mantle-to-surface ratio, \textit{nM/S}. It allows one to avoid the effects caused by different proportion of various elements in the ice phases and different distribution of the total number of metal atoms among the mantle and surface phases, both of which can vary over the integration period. The full calculation formula for \textit{nM/S} is
\begin{equation}
   \label{ms2}
nM/S_i=\frac{n_{i,M} \times \sum^{I}_{J}a}{n_{i,M}\sum^{I}_{J}(a \times R_E)},
   \end{equation}
where \textit{a} (the index in a molecular formula) is the number of atoms of elements \textit{I} to \textit{J} in the molecule \textit{i}. $R_E$ is the total \textit{M/S} ratio for each element that is in the molecule \textit{i} (C, N, O, S from \textit{I} through \textit{J}). Na, Mg, and Fe do not form molecules in the model. Because hydrogen wanders rather freely between the surface and the mantle, it is not counted here. $R_E$ for an element is calculated by
   \begin{equation}
   \label{ms3}
R_{E}=\frac{\sum^{i}_{j}(m \times n_{i,M})}{\sum^{h}_{j}(m \times n_{i,S})},
   \end{equation}
i.e., $R_E$ is the total relative abundance of the element \textit{E} over all species \textit{i} through \textit{j} in the mantle-phase divided by the total relative abundance of this element in the surface-phase.
   \begin{figure}
	 \vspace{2.5cm}
	 \hspace{-1.0cm}
   \includegraphics[width=15.0cm]{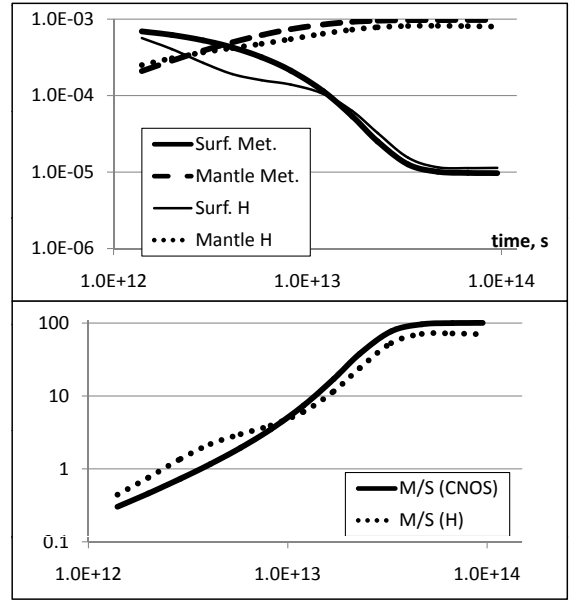}
	 \vspace{-6.0cm}
   \caption{
Top: evolution of the total relative abundance for the mantle-forming metal elements C, N, O, S, and for hydrogen, chemically bound in metal molecules. Bottom: evolution of the \textit{M/S} ratio for metals and chemically bound H.}
   \label{graf-str}
   \end{figure}

\subsection{Structure of the mantle}
\label{res-str}

The structure of the mantle (surface roughness, porosity, except for closed cavities) is characterized by the \textit{M/S} ratio for the total abundance of metal atoms in the mantle and surface phases. Fig.~\ref{graf-str} shows the abundance and $M/S_E$ for chemically active ice metal elements and chemically bound H in ice. The enhancement of the cumulative \textit{M/S}$_{CNOS}$ reflects the compaction of the mantle over time, i.e., the reduction of the abundance of molecules in the outer surface phase.

One can conclude from the data of Fig.~\ref{graf-str} that an approximate state of equilibrium in the mantle-forming process is reached in the last two time-steps only, when the differences between \textit{M/S}$_{CNOS,t}$ have become smaller than 1 \%. At $t=6.7 \times 10^{13}$ s \textit{M/S}$_{CNOS}=99.9$, at $t=9.5 \times 10^{13}$s \textit{M/S}$_{CNOS}=100.5$ , and at $t = 10^{15}$ s, which is another 6-7 time steps forward, \textit{M/S}$_{CNOS}= 101.0$. The corresponding metal total relative abundances $n_{M,CNOS}$ and $n_{S,CNOS}$ change only slightly, below 1 \% if $t \geq 6.7 \times 10^{13}$ s.

\subsection{General results}
\label{res-gen}
   \begin{figure*}
	 \vspace{5.0cm}
	 \hspace{-1.5cm}
   \includegraphics[width=36.0cm]{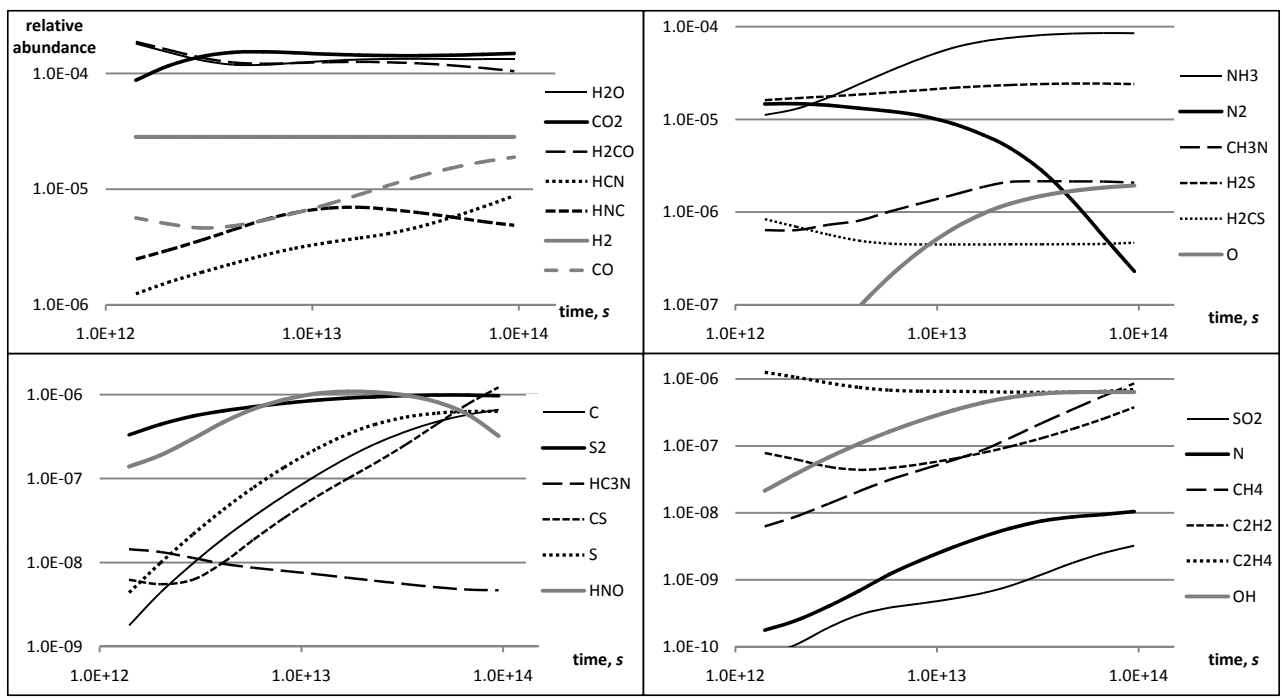}
   \vspace{-17.5cm}
   \caption{Evolution of the solid-phase relative abundance (surface + mantle) for selected species. Note the scale of each graph.}
   \label{graf-gen-ab}
   \end{figure*}
   \begin{figure*}
	 \vspace{5.0cm}
	 \hspace{-1.0cm}
   \includegraphics[width=28.0cm]{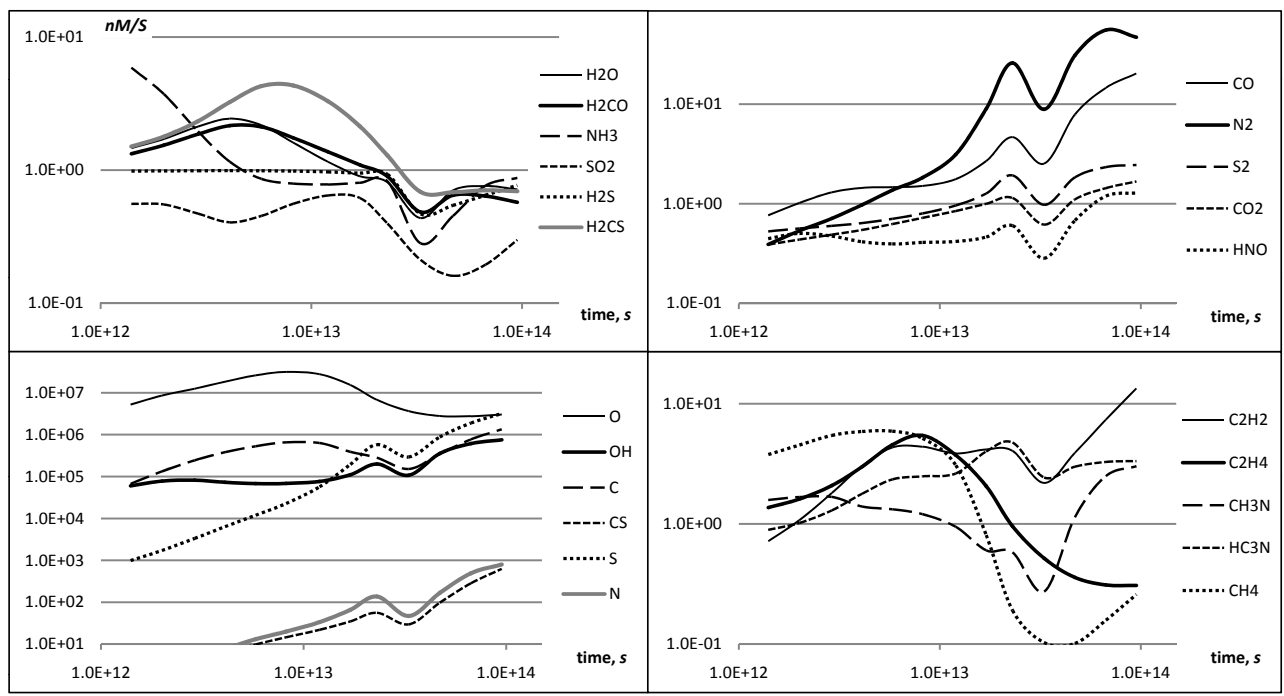}
	 \vspace{-17.5cm}
   \caption{Evolution of the normalized mantle-to-surface \textit{nM/S} ratio for selected species. Note the scale of each graph.}
   \label{graf-gen-nms}
   \end{figure*}
This section includes the analysis of the calculated abundances for surface and mantle phase species. Figures~\ref{graf-gen-ab} and~\ref{graf-gen-nms} show the relative abundance of species in ice phases and the \textit{nM/S} ratio, respectively. These are the main results of the calculations.

The most important species in ice phase are small molecules – H$_2$O, CO$_2$, and H$_2$CO, and also NH$_3$, H$_2$S, CO, and N$_2$. The total proportion of single atoms and other radicals (O, C, S, OH, NH$_2$, NH, etc.) is lower than 1 \%, and they contain an even lower proportion of metal atoms. A growing abundance curve and high \textit{nM/S} is characteristic of radicals such as free atoms, OH, SO, CN, CH$_2$, C$_2$H, etc.

The mantle has a lower hydrogen content, which indicates that the outward hydrogen diffusion plays a more important role than inward diffusion. When the integration time becomes longer than approximately $10^{13}$s, \textit{M/S}$_{\mathrm{H}} \approx 0,7M/S_{\mathrm{CNOS}}$ (compare Fig.~\ref{graf-str}). This means that the hydrogen content bound in metal molecules in the mantle is only around 70 \% of that in surface molecules. This is mostly because
 \textit{nM/S} < 1 
for the important H$_2$O, H$_2$CO, NH$_3$, and H$_2$S molecules. The oxygen disposed by the reduction of H$_2$O abundance ends up as part of CO$_2$. This is a significant result since it testifies to the capability of the novelties in the model to address an important astrochemical problem – the route of synthesis of CO$_2$ in interstellar ice. It is produced in subsurface processes, where CO is oxidized by the products of H$_2$O photodissociation, mainly O \citep[see also][]{Ioppolo09, Ioppolo11, Oba10a, Oba10b, Roser01}. The CO conversion in H$_2$O matrix may address the difficulty in observing features related to CO:H$_2$O mixture in interstellar ices \citep{Pontoppidan03}.

Carbon monoxide is the main gas-phase molecule and is the main form of C and O accreting onto grain surface at conditions given \citep{Kalvans12}. On the surface it can attach itself to either H or O and transform into H$_2$CO or CO$_2$. Formaldehyde is the most abundant form of hydrogenated CO. The relative abundance of methanol, which is an abundant hydrogenation product of CO in interstellar ice \citep{Oberg11-109, Peng12}, is negligible. This is because the data from \citeauthor{Hasegawa92} lack the hydrogenation reactions of formaldehyde.

OCS is the only ice-phase, sulfur-bearing molecule that has been observed in space \citep{Gibb04, Aikawa12}. It is among the molecules with highest \textit{nM/S}. Other molecules whose abundance in the mantle is much higher than the abundance in the surface
(\textit{nM/S} > 100)
are CN, HCS, carbon chains, and smaller radical-type compounds (mobile species). These species have been directly or indirectly created by the dissociation by CR-photons of stable molecules. CO has the highest \textit{nM/S} of the stable major molecules, because it is also the dissociation product of H$_2$CO and CO$_2$. The production of carbon-chain compounds in ice complies with the observations of protostars \citep{Bottinelli04, Kuan04}.

The general results of this model (less H and H$_2$O, CO$_2$ as the dominating species in the ice) agree with the results in \citetalias{Kalvans10}. They also approximately agree with observations \citep[e.g.][]{Oberg11-109}. The previously calculated distribution of sulfur compounds in the ice (with a similar abundance for H$_2$S, SO$_2$, SO, H$_2$CS, and CS) better agrees with observations \citep{Hatchell98, Tak03, Wakelam04, Herpin09, Ren11} than the current results with H$_2$S as the single most important sulfur molecule. This and the calculation results at $t = 10^{15}$ s (Sect.~\ref{res-1e15}) may indicate that the ice is more intensively processed before the protostar evaporates the mantle. The cause can be the chemical changes induced by cosmic rays, and the heating of ice in young stellar objects (YSO).

The main difference between the composition of the surface and the mantle is the amount of CO hydrogenation products. The relative abundance of H$_2$CO on the surface surpasses even that of H$_2$O, while in the mantle it is significantly lower and complies better with reality. This can be an evidence for the usefulness of subsurface processes in explaining the composition of interstellar ices. The total calculated abundance of hydrogenated CO still significantly exceeds the observed H$_2$O:CH$_3$OH 100:4 ratio. This is probably due to the limited reaction set and the absence of any measure to take into account the discrete nature of the reactions on grains (modified rate-equation or Monte-Carlo approach, e.g. \citet{Charnley01,Stantcheva01,Stantcheva04}). In real ices some of these molecules are turned into complex organic molecules \citep{Oberg11-14}.

\subsection{Results and discussion on deuterium species}
\label{res-d}
   \begin{figure*}
	 \vspace{5.0cm}
	 \hspace{-1.5cm}
   \includegraphics[width=27.5cm]{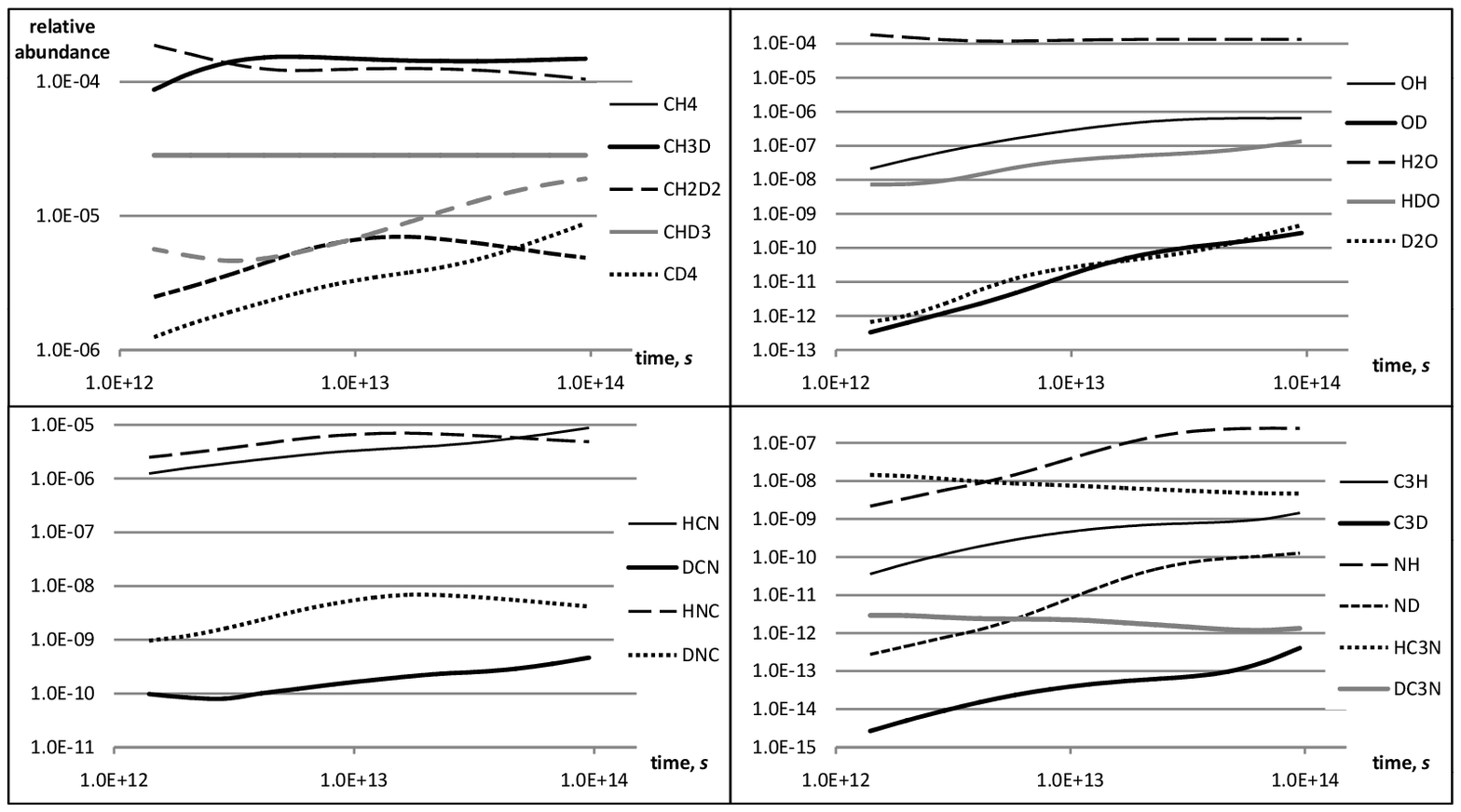}
	 \vspace{-17.0cm}
   \caption{
Examples of calculation results on the abundance of deuterated species. The evolution of the solid-phase (surface + mantle) relative abundance of selected hydrogen-isotopologs is shown. Note to the scale of each graph.
 			}
    \label{graf-d-ab}
    \end{figure*}
%
   \begin{figure*}
	 \vspace{5.0cm}
	 \hspace{-1.5cm}
   \includegraphics[width=27.0cm]{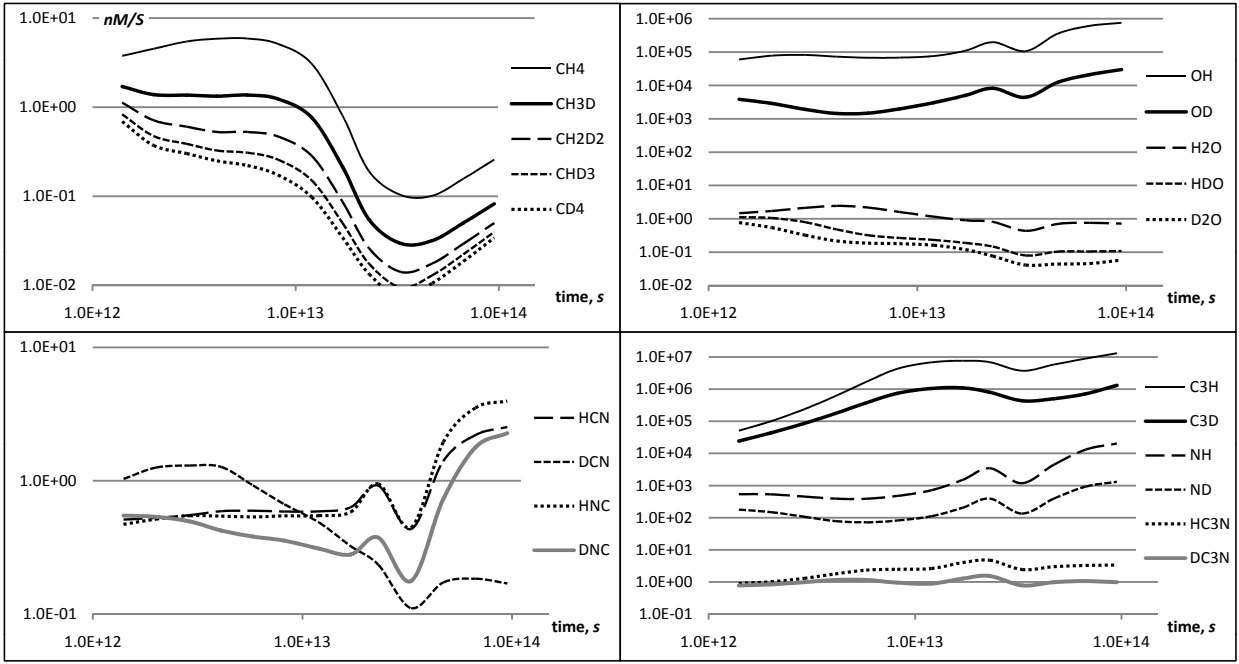}
	 \vspace{-16.5cm}
   \caption{
Examples of calculation results on the \textit{nM/S} ratio for deuterated species. The evolution of the \textit{nM/S} for all D isotopologs of selected species is shown.  Note the scale of each graph.
 			}
    \label{graf-d-nms}
    \end{figure*}
%
   \begin{figure*}
	 \vspace{5.0cm}
	 \hspace{-1.5cm}
   \includegraphics[width=27.0cm]{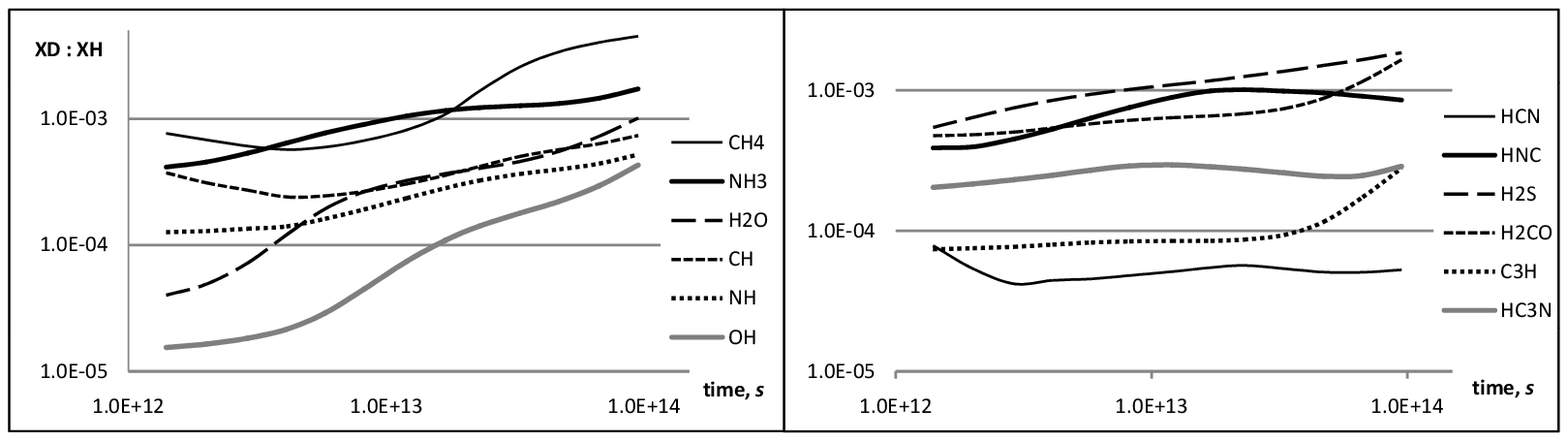}
	 \vspace{-21.0cm}
   \caption{
Calculated deuteration level for selected species. The evolution of the abundance ratio ($R_\mathrm{D}$) for singly deuterated to non-deuterated molecules in the solid phase (mantle + surface) is shown.
 			}
   \label{graf-d-rd}
   \end{figure*}

\begin{figure}
	 \vspace{2.5cm}
	 \hspace{-1.0cm}
   \includegraphics[width=15.0cm]{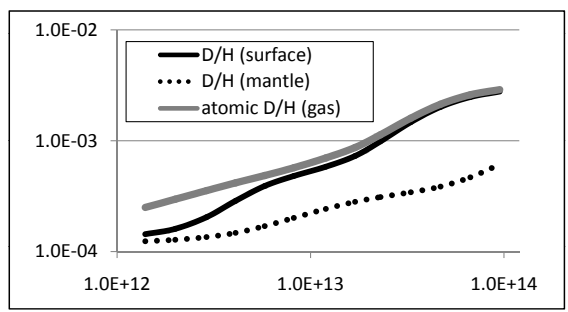}
	 \vspace{-10.0cm}
   \caption{
Evolution of the chemically bound deuterium to chemically bound hydrogen total abundance ratio in the solid phases and the gas-phase atomic D-to-H abundance ratio.
			}
   \label{graf-d-hd}
   \end{figure}

The calculation results regarding the molecule enrichment in deuterium are shown in four figures.  Fig.~\ref{graf-d-ab} shows the total abundance in ice (mantle + surface phases) for selected species.  The \textit{nM/S} ratio is shown in Fig.~\ref{graf-d-nms} and the deuteration ratio in Fig.~\ref{graf-d-rd}. The deuteration ratio or $R_\mathrm{D}$ is the abundance ratio XD : XH, where X is the remainder of a molecule with H or D atom. Fig.~\ref{graf-d-hd} shows the total abundance ratio for chemically bound H and D in surface- and mantle phases.

The hydrogen-containing heavy molecules can be divided into three broad classes. First, there are those that are intensively produced in the mantle. These include HCN, C$_3$H, C$_4$H, H$_2$S$_2$, HCOOH, and others. At the first stage they are produced with some level of deuteration on the surface, then they are buried and become part of the mantle. There they are produced additionally, with a deuteration characteristic of the mantle and their level of deuteration may even not grow with time, unlike all other species.

The second group of molecules (e.g. HNO, C$_2$H$_4$, CH$_4$, H$_2$CO, NH$_3$) are mostly destroyed, not produced in the mantle, so they keep the comparatively high level of deuteration achieved by their synthesis in the gas phase and on the outer surface, and with time it only becomes higher.

Between these two groups are species with \textit{nM/S} values around unity, most notably H$_2$O as the most abundant hydrogenated heavy molecule. The abundance, \textit{nM/S}, and deuteration level for examples of all three groups are shown in Figs.~\ref{graf-d-ab},~\ref{graf-d-nms}, and~\ref{graf-d-rd}.

Fig.~\ref{graf-d-hd} shows that the D enrichment in surface molecules is much higher than in the mantle. For both surface and mantle, $R_\mathrm{D}$ is higher than the cosmic D/H ratio of $2 \times 10^{-5}$. Generally, modeling results indicate that the mantle molecules lose a significant proportion of D when they are included in the mantle. $R_\mathrm{D}$ grows with time in the solid phases. For most cases, the relative abundance of D-containing molecules correlates with their protium analogs.

Free hydrogen atoms have $R_\mathrm{D}$ > 200 at $t \approx 3$ Myr, it is the highest of all species in the mantle. Metal molecules with highest $R_\mathrm{D}$ are organic species with a rather high hydrogen content (CH$_4$, C$_2$H$_4$, CH$_2$NH, CH$_3$CN, CH$_3$OH, H$_2$CO, etc.) and H$_2$S, NH$_3$. Species with lowest $R_\mathrm{D}$ are radicals, HCN, H$_2$S$_2$, and organic species with low hydrogen content (HC$_3$N, C$_4$H, C$_3$H, C$_2$H, and HCOOH, see Fig.~\ref{graf-d-rd}).

The $R_\mathrm{D}$ for gas-phase species is similar to the surface molecules, it is 1-3 times higher than the $R_\mathrm{D}$ for solid phase (total) species. The observed pattern of D enrichment can be explained assuming that the maximum $R_\mathrm{D}$ is reached in reactions in the gas-phase. The molecules on the surface are in direct contact with molecules accreting from the gas and are cycled through gas- and surface phases by desorption and accretion processes. They mostly keep their high $R_\mathrm{D}$. When they are incorporated into the mantle, they lose connection to the efficient gas-phase deuterium enrichment ``factory''. Their $R_\mathrm{D}$ is determined by the proportion of H and D that is incorporated into ice, with atomic hydrogen as the most important form. D/H in the gas-phase (also shown in Fig.~\ref{graf-d-hd}) is typically lower than $R_\mathrm{D}$ for heavy molecules, so ice molecules lose their high deuteration ratios inherited from gas-phase chemistry. This mechanism can also serve as a partial explanation of selective D enrichment in ice compounds, because molecules hydrogenated in different phases have different $R_\mathrm{D}$.
%

The most important hydrogen-containing molecule is water. Its calculated final abundance in ice is similar to the observed one, with the HDO content a few times $10^{-4}$ \citep{Peng12} and [D$_2$O]/[HDO] $10^{-3}$ \citep{Butner07}. The $R_\mathrm{D}$ for formaldehyde are higher than those of H$_2$O (as expected) but lower than the values from observations \citep[see e.g.][Table 1 and references therein]{Cazaux11}. Formaldehyde is overproduced in the model. The average $R_\mathrm{D}$ is approximately consistent with D enrichment in starless prestellar core, which is the main subject of this paper. For most other molecules the maximum calculated $R_\mathrm{D}$ is 0.5 \% or less. $R_{\mathrm{D}}$ is far from the observed maximum enrichment of several per cent \citep{Cazaux11, Coutens12}. Observed values exceed 10 \%, especially for molecules with several hydrogen atoms (up to 30 \% CH$_2$DOH for methanol, \citet{Parise02}). The highest $R_\mathrm{D}$ is observed in protostars.

The calculated $R_\mathrm{D}$ and the dynamics of enrichment growing over time generally agree with observations of molecular gas and YSOs. However, the modeling results are different from the existing view that interstellar ice molecules are always highly enriched in D. Such a product can be an artifact due to the approximations used in the model. For example, the cause may be the difference in the photodissociation rate and reaction branching ratios between H and D species.

If the effect is real, it may place some constraints on the interpretation of PSC, YSO and protoplanetary disk observations. This product is different from our previously published results \citep{Kalvans11} because of a different and more cautious approach to the description of the diffusion of H and D. The overall modeling results, including those at $t=10^{15}$ s, point to a possible measurable delay in the D-enrichment process of ice molecules. The delay might be shorter than that obtained from model results (at least several Myr), because high $R_\mathrm{D}$ can be reached more rapidly due to the chemical changes induced directly by cosmic rays.

\subsection{Calculation results and discussion for a model at $t=10^{15}$s}
\label{res-1e15}
\begin{table}
\caption{The parameters for the content of chemically bound H and D in the mantle and surface phases for integration times $9.5 \times 10^{13}$s and $10^{15}$s. \textit{n} is the relative abundance, chem. -- chemically bound H or D in metal molecules.}
\label{tab-1e15}
	\centering
	\small
		\begin{tabular}{lcc}
\hline\hline
Parameter&$t=9.5^{13}$ s&$t=1.0^{15}$ s\\
\hline
\textit{n}, free H in mantle&2.43E-12&3.67E-12\\
\textit{n}, free D in mantle&6.15E-10&3.31E-09\\
\textit{n}, free H$_2$ in mantle&2.83E-05&2.83E-05\\
\textit{n}, free HD in mantle&9.05E-10&8.64E-10\\
\textit{n}, free D$_2$ in mantle&9.11E-13&1.05E-12\\
\textit{n}, chem. H, surface&1.14E-05&1.19E-05\\
\textit{n}, chem. H, mantle&7.98E-04&8.60E-04\\
\textit{M/S} for chem. H&7.02E+01&7.22E+01\\
chem. D:H, mantle ($R_{D,M+C}$)&6.10E-04&2.29E-03\\
chem. D:H, surface ($R_{D,S}$)&2.77E-03&3.12E-03\\
atomic [D]/[H], gas &2.89E-03&3.14E-03\\
\hline
		\end{tabular}
\end{table}
There are indications from the modeling results that the interstellar ice has been more intensively processed than assumed in the model. Because of this, and also to bridge the integration time gap between this work and \citetalias{Kalvans10}, calculations with a prolonged integration time of $10^{15}$s were performed. They are also relevant to long-lived ices in non-starforming cores. The results obtained are different from results with shorter \textit{t} and interesting enough to be briefly analyzed. To facilitate the interpretation, these results are compared with results with $t=9.5 \times 10^{13} \approx 10^{14}$s in Table.~\ref{tab-1e15}.

The total \textit{M/S} ratio for both simulations is similar (Sect.~\ref{res-str}), but the content of H and D in mantle molecules differs (Table~\ref{tab-1e15}). At $t=10^{15}$ s the content of chemically bound H in subsurface molecules has increased by ~8 \%, and by 4 \% for surface species. This is reflected in the abundance of major species, because the most abundant molecule is now H$_2$O, not CO$_2$. The HCN:HNC ratio is still higher, the abundance of N$_2$ has increased, and H$_2$CO, CH$_3$OH have diminished. These changes affect important molecules and are more consistent with observations. The abundance has grown considerably (more than five times) for several species with different \textit{nM/S}. This suggests that the enhanced production is phase-independent and is a result of the reaction net, where the molecule is involved. The abundance of atoms and other free radicals has remained approximately the same.

The next important result for the $t=10^{15}$s model is the increase of the proportion of chemically bound D. The total $R_\mathrm{D}$ for mantle species is now 2/3 of the surface $R_\mathrm{D}$. The gas-phase atomic D/H ratio, which determines $R_\mathrm{D}$ for subsurface ice species, is growing over time. The average $R_\mathrm{D}$ for mantle molecules is growing with a delay, because of the inefficient processing of the mantle.

At $10^{15}$ s, $R_\mathrm{D}$ has increased for virtually all molecules, except H$_2$. For important species, such as H$_2$O, H$_2$CO, and NH$_3$, this increase is four times. Like the changes in species’ abundance, the enhancement of $R_\mathrm{D}$ is independent from the \textit{nM/S} ratio. Moreover, changes in D enrichment are practically independent of the abundance changes in results at $t=10^{14}$ and $10^{15}$s.

Generally, the results with $t=10^{15}$s make the impression that the ice on grains is perhaps subjected to a more prolonged or more intense chemical and physical processing than permitted by the model, because they are more consistent with observations.

\section{Discussion}
\label{disc}

The model used for calculations includes a full system that is able to at least partially describe subsurface processes in the mantle. It is rather general and superficial in terms of the reactant and reaction databases. It seldom reproduces abundances of specific molecules that are consistent with observations. The approach used was to analyze the differences produced by subsurface chemistry relative to the ``conventional'' chemistry on the outer surface. In addition to the results described in Sect.~\ref{results}, it allowed us to draw some more general conclusions regarding the chemistry of interstellar ice.

The developed model allows proposing an interpretation of the processes that occur in ice with two chemically different layers \citep{Schutte91}. The inner layer (closer to the grain nucleus) consists of H$_2$O, CO$_2$, NH$_3$, and other molecules, mostly characterized by the ability to form hydrogen bonds. Volatile species, such as CO, CH$_4$, and N$_2$ form the outer layer. This is more exposed to radiation and absorbs a larger dose of UV photons that tear down its molecules. However, thanks to the weaker binding between molecules, the radicals are more mobile and able to recombine effectively (i.e., in terms of our model, its cavities would resurface more frequently). The layer underneath is more shielded and fewer radicals are produced there, while its sturdier, hydrogen-bound structure is less affected by Fe-CR hits and other energetic events. This picture improves the current view on the processes in ice layers \citep[see, e.g.][]{Linnartz11}.

Our work allows us to suggest that desorption by chemical explosions \citep{Greenberg73, dHendecourt82, Schutte91} is perhaps insignificant in interstellar clouds. In an explosion the grain loses whole segments of ice, which leads to the destruction of the dense, layered mantle structure. At 10 K or at an even lower temperature the ice formed by accretion can be expected to be porous. Ice segregation and compaction is a long-term process and no ice layers of different chemical composition will be able to form if repeated explosions occur. Instead, the individual reactions in ice slowly decrease the excess abundance of radicals, and are an important feature in explaining the observed ice composition. Additionally, according to our model, a long lifetime of ice is required to explain the high observed D enrichment in molecules.

Atomic hydrogen may provide an important contribution to the reactions producing an ice explosion \citep{Rawlings13}. However, according to our calculation results, during astronomical timescales H will diffuse from the mantle to the surface before evaporating to the gas phase, i.e., it does not accumulate in the ice mantle and cannot take part in explosive processes. $D_\mathrm{H}$ is the most trustworthy of all diffusion coefficients for light hydrogen species used in this paper. It was obtained by \citet{Awad05} from the experiments by \citet{Watanabe03} with interstellar ice analogs, H$_2$O:CO mixture. Yet no significant accumulation of H in the mantle was observed in the calculation results (Table~\ref{tab-phases}). This further decreases the likelihood for chemical explosions to play a significant role in processes regarding interstellar ices.

Calculation results regarding the deuterium enrichment in molecules (Sect.~\ref{res-d}) predict that there might be a prolonged period in the evolution of starless and star-forming cores, when the enrichment in ice molecules is low. It is possible that this period is shorter than the results indicate (around 10 Myr) because of the additional processing by cosmic rays. The conclusion that in quiescent cloud cores $R_\mathrm{D}$ for molecules in the ice mantles is lower than in matter affected by the formation of a protostar agrees with observations, but there are other possible explanations, too \citep{Parise05, Bergman11}.

\begin{acknowledgements}
The publication of this article is financed by ERDF project SATTEH, No. 2010/0189/2DP/2.1.1.2.0/10/APIA/VIAA/019, being implemented in Engineering Research Institute ``Ventspils International Radio Astronomy Center'' of Ventspils University College (ERI VIRAC).

The authors are thankful to Institute of Astronomy, University of Latvia, for providing technical support for the calculations performed for the article. We thank the anonymous referee for many valuable comments and suggestions. This research has made use of NASA's Astrophysics Data System.
\end{acknowledgements}

\bibliographystyle{aa}
\bibliography{dmantle}

\end{document}